\documentclass[prc,twocolumn,showpacs,nofootinbib,preprintnumbers]{revtex4-1}

\usepackage{hyperref}
\usepackage{graphicx}
\usepackage{bm}
\usepackage{amsmath}

\begin{document}
\preprint{YITP-16-6}

\title{Probing neutron-proton dynamics by pions}

\author{Natsumi Ikeno}
\affiliation{Department of Regional Environment, Tottori University,
  Tottori 680-8551, Japan}
\author{Akira Ono}
\affiliation{Department of Physics, Tohoku University,
  Sendai 980-8578, Japan}
\author{Yasushi Nara}
\affiliation{Akita International University,
  Akita 010-1292, Japan}
\author{Akira Ohnishi}
\affiliation{Yukawa Institute for Theoretical Physics, Kyoto University,
  Kyoto 606-8502, Japan}

\begin{abstract}
In order to investigate the nuclear symmetry energy at high density,
we study the pion production in central collisions of neutron-rich
 nuclei ${}^{132}\mathrm{Sn}+{}^{124}\mathrm{Sn}$ at 300 MeV/nucleon
using a new approach by combining the antisymmetrized
molecular dynamics (AMD) and a hadronic cascade model (JAM).  
The dynamics of neutrons and protons is solved by AMD, and then pions
 and $\Delta$ resonances in the reaction process are handled by JAM.
 We see the mechanism how the $\Delta$ resonance and pions are
 produced reflecting the dynamics of neutrons and protons.  We also
 investigate the impacts of cluster correlations as well as of the
 high-density symmetry energy on the nucleon dynamics and consequently
 on the pion ratio.  We find that the $\Delta^-/\Delta^{++}$
 production ratio agrees very well with the neutron-proton squared
 ratio $(N/Z)^2$ in the high-density and high-momentum region.  We
 show quantitatively that $\Delta$ production ratio, and therefore
 $(N/Z)^2$, are directly reflected in the $\pi^-/\pi^+$ ratio, with
 modification in the final stage of the reaction.
\end{abstract}

\maketitle
\section{Introduction}
It is one of the important subjects in nuclear physics and
astrophysics to determine the nuclear symmetry energy at various
densities.  Constraints on the symmetry energy at densities lower than
the normal density $\rho_0$ have been obtained to some degree from
nuclear physics experiments \cite{Tsang:2012,Horowitz:2014}.  On the
other hand, intermediate-energy heavy-ion collisions are believed to
be useful to investigate the symmetry energy at high densities.  In
fact, in central collisions at several hundred MeV/nucleon, transport
calculations show that the maximum density around $2\rho_0$ is reached
when the system is compressed in an early stage of the reaction.  The
ratio of the neutron and proton densities in the compressed part is
naturally sensitive to the symmetry energy at high densities
\cite{BaoAnLi:2008,Li}.  However, it is necessary to
understand the link between the effects in the compression stage and
the final observables, in order to extract the high-density symmetry
energy from experimental data available at present and in future.

The $\pi^{-}/\pi^{+}$ ratio has been proposed to be a good probe to
constrain the high-density behavior of the symmetry energy~\cite{Li}.
In heavy-ion collisions, the pions are produced through the $\Delta$
resonance formation in the nucleon-nucleon collisions that typically
occur at early times in the compressed part of the system.  If all the
pions produced by the $NN\rightarrow N\Delta\rightarrow NN\pi$ process
are directly emitted, the expected ratio should be
$\pi^-/\pi^+=(5N^2+NZ)/(5Z^2+NZ)\approx(N/Z)^2$, where $N$ and $Z$ are
the numbers of neutrons and protons that are relevant for the $\Delta$
production.  In another case of chemical equilibrium for
$NN\leftrightarrow N\Delta$ and $\Delta\leftrightarrow N\pi$ reactions
at a temperature $T$, the ratio $\pi^-/\pi^+ \simeq e^{2(\mu_n-\mu_p)/T}$
\cite{Bertsch:1980,Bonasera:1987}, where $\mu_n$ and $\mu_p$ are the
neutron and proton chemical potentials, may also be related to the
neutron and proton densities $\rho_n/\rho_p$ or the symmetry energy
\cite{Li,Ferini:2005}.  We also notice the relation
$e^{2(\mu_n-\mu_p)/T}\approx[f_n(\varepsilon)/f_p(\varepsilon)]^2$
when the neutron and proton phase-space densities are compared at the
same single-particle energy $\varepsilon$ satisfying
$\varepsilon-\mu_{n,p}\gg T$.  Thus, both in these extreme cases, the
$\pi^-/\pi^+$ ratio is related to some kind of neutron-to-proton
squared ratio $(N/Z)^2$ which is then supposed to be sensitive to the
symmetry energy at high densities.

Some theoretical studies have been performed by different transport
models to investigate the sensitivity of pion
observables~\cite{Li,Gaitanos:2004,Reisdorf:2007,IBUU04,ImIQMD,pBUU,BUU}.
At present, however, some of these results are contradicting to each
other even qualitatively.  The symmetry-energy dependence of the
$\pi^-/\pi^+$ ratio in Ref.\ \cite{ImIQMD} is opposite to that
predicted by Refs.\ \cite{Li,IBUU04}, while Ref.\ \cite{pBUU} predicts
that the $\pi^-/\pi^+$ ratio for the total pion multiplicities does not
depend on the density dependence of the symmetry energy.  Moreover,
the $\pi^{-}/\pi^{+}$ ratio in central Au + Au collisions measured by
the FOPI Collaboration~\cite{Reisdorf:2007,FOPI} is actually larger
than the squared neutron-to-proton ratio of the total system, 
$(N/Z)^2_{\text{sys}}$, at low energies such
as at 400 MeV/nucleon.  Since the compressed part naturally becomes
less neutron rich than the total system, we inevitably have
$\pi^-/\pi^+>(N/Z)^2_{\text{sys}}>(N/Z)^2$, and therefore
$\pi^{-}/\pi^{+}$ cannot agree with $(N/Z)^2$, contradicting to the
arguments in the previous paragraph, as long as $N/Z$ is identified
with the density ratio $\rho_n/\rho_p$ in the compressed part as is
typically done in the literature.  Most of the transport calculations
underestimate the measured pion ratio, but some recent calculations
\cite{ImIQMD,pBUU} predict the ratio as high as the experimental data.
Nevertheless, investigations cannot be found in the literature on the
mechanism to increase the pion ratio.  Hence, our understanding of the
pion ratio in relation to the symmetry energy and the nucleon dynamics
is not complete particularly when we need to have a precise
description to explain the absolute magnitude of $\pi^-/\pi^+$.

In this paper, we study the pion production in central collisions of
neutron-rich nuclei ${}^{132}\mathrm{Sn}+{}^{124}\mathrm{Sn}$ at 300
MeV/nucleon, which is one of the systems to be measured at RIBF/RIKEN~\cite{RIBF}.
We develop and employ a new approach by combining the antisymmetrized
molecular dynamics (AMD)~\cite{AMD} and a hadronic cascade model
(JAM)~\cite{JAM}.  The dynamics of neutrons and protons is solved by
AMD, and then pions and $\Delta$ resonances in the reaction process
are handled by JAM.  AMD calculations were performed for several cases
with and without cluster correlations \cite{Ono:NN2012}, and with two
effective interactions corresponding to different density dependences
of the symmetry energy.  These different cases of AMD calculation
yield different dynamics of neutrons and protons.  The main aim here
is to utilize these different cases to identify how the $\Delta$
resonances and the emitted pions carry the information of the nucleon
dynamics.  We will also see the impacts of cluster correlations as
well as of the high-density symmetry energy on the nucleon dynamics
and consequently on the pion ratio.

\section{Formulation}
In order to understand the $\pi^-/\pi^+$ enhancement mechanism
and to extract high-density symmetry energy, we need a reliable transport
model of nucleons, clusters, $\Delta$ resonances and pions.
AMD has been demonstrated to be a reliable transport model of nucleon and
clusters~\cite{AMD,Ono:NN2012}.
It takes account of nucleon mean field effects, two-nucleon elastic
scatterings,
event-by-event fluctuations and antisymmetrization of nucleon wave
functions.
All of these ingredients could affect the $N/Z$ ratio during the collision,
and thus have influences on the $\pi^-/\pi^+$ ratio.
One can explain fragment mass distribution in heavy-ion collisions
precisely,
especially with cluster correlations.
In the present AMD code, however, $\Delta$ resonances and pions have not
been incorporated. 
On the other hand,
JAM is a reliable hadron transport model~\cite{JAM}.
It has been successfully applied to $pA$ and $AA$ collisions in the energy
range
from 1 GeV/nucleon to 158 GeV/nucleon.
We can describe hadron production spectra in JAM,
and collective flows are also explained with the mean field effects
switched on \cite{Isse:2005nk}.
By comparison, antisymmetrization of nucleon wave functions is not included,
and cluster correlation during the time evolution are not implemented.
At present, there is no transport model which includes all of the above
mentioned
important ingredients;
mean field effects, two-body collisions including elastic and inelastic
processes,
event-by-event fluctuations, antisymmetrization of nucleon wave functions,
and dynamical cluster correlations.
Thus one of the best ways would be to combine AMD and JAM
--- nucleon transport in AMD and particle production in JAM.

\subsection{Perturbative treatment of pions and $\Delta$}

We consider here a transport model which describes the reaction
dynamics by the time evolution of one-body density matrices or
corresponding phase-space distributions $f_\alpha(\bm{r},\bm{p},t)$,
where the index $\alpha$ stands for the particle species such as
nucleons ($N$), $\Delta$ resonances ($\Delta$) and pions ($\pi$).  It
should be implicitly understood that the isospin (and spin) components
of these particles are also distinguished by $\alpha$.  The coupled
equations for $f_\alpha$ ($\alpha=N,\Delta,\pi$) can be written in
general as
\begin{equation}
\frac{\partial f_\alpha}{\partial t}
= \dot{f}^{\text{MF}}_\alpha[f_N,f_\Delta,f_\pi]+I_\alpha[f_N,f_\Delta,f_\pi]
\end{equation}
with the mean-field term $\dot{f}^{\text{MF}}$ and the collision term
$I_\alpha$, both of which depend on the phase-space distributions of
all the particles at the time $t$ in general.  The collision term
includes at least the following processes
\begin{gather}
N+N\rightarrow N+N,\\
N+N\rightarrow N+\Delta,\\ N+\Delta\rightarrow N+N,\\
\Delta\rightarrow N+\pi,\\ N+\pi\rightarrow\Delta.
\end{gather}

Now we limit our consideration to the sub-threshold or near-threshold
cases where only a small number of $\Delta$ production processes occur
because the incident energy is not so high.  Then we expect that the
$\Delta$ production can be treated perturbatively.  If the
perturbation parameter $\lambda$ is multiplied to the probability of
the $NN\rightarrow N\Delta$ process, $\Delta$ resonances and pions
will appear in the first order of $\lambda$,
$f_\Delta=\lambda f_\Delta^{(1)}+O(\lambda^2)$ and
$f_\pi=\lambda f_\pi^{(1)}+O(\lambda^2)$.  The zeroth order equation
for the nucleon distribution $f_N=f_N^{(0)}+O(\lambda)$ is
\begin{equation}\label{eq:fn0}
\frac{\partial f_N^{(0)}}{\partial t}
=\dot{f}^{\text{MF}}_N[f_N^{(0)},0,0]+I_N^{(\lambda=0)}[f_N^{(0)},0,0],
\end{equation}
where the collision term $I_N^{(\lambda=0)}$ includes only the elastic
$NN\rightarrow NN$ scatterings.  The distributions of $\Delta$
resonances and pions can be solved
by
\begin{align}
\frac{\partial f_\Delta}{\partial t}
&=\dot{f}^{\text{MF}}_\Delta[f_N^{(0)},f_\Delta,f_\pi]
+I_\Delta[f_N^{(0)},f_\Delta,f_\pi],
\label{eq:fdelta}
\\
\frac{\partial f_\pi}{\partial t}
&=\dot{f}^{\text{MF}}_\pi[f_N^{(0)},f_\Delta,f_\pi]
+I_\pi[f_N^{(0)},f_\Delta,f_\pi],
\label{eq:fpi}
\end{align}
which are correct up to the first order of $\lambda$.

The equation (\ref{eq:fn0}) for the zeroth order of the nucleon
distribution can be solved assuming a system composed of only nucleons
without considering productions of other particles.  In our present
approach, we solve the nucleon dynamics by AMD \cite{AMD,Ono:NN2012}.
Then, for the calculated $f_N^{(0)}$, the equations (\ref{eq:fdelta})
and (\ref{eq:fpi}) for $\Delta$ resonances and pions are solved by
another transport model JAM \cite{JAM} which can handle particle
productions.  The information of nucleons in the JAM calculation is
always replaced by $f_N^{(0)}$ calculated by AMD.  Namely the particle
production is calculated by JAM based on the nucleon dynamics
calculated by AMD.

The above treatment violates some conservation laws in the higher
orders $O(\lambda^2)$.  The result may be improved by introducing
corrections for the conservation laws of baryon number, change and
energy, modifying the nucleon information $f_N^{(0)}$ in
Eqs.~(\ref{eq:fdelta}) and (\ref{eq:fpi}).  When a $\Delta$ resonance
exists in the JAM calculation, it should have been produced by a
collision of two nucleons.  Such a pair of two nucleons is chosen in
the AMD calculation by taking into account the distance from the
$\Delta$ resonance, the phase space to produce a $\Delta$ resonance,
and the charge conservation condition.  Then one of the two nucleons
is annihilated (assuming that it is replaced by the $\Delta$
resonance) and the charge and the momentum of the other nucleon is
modified for the charge and energy conservations.  In the pion case,
we consider only the charge conservation by modifying the charge of a
nucleon for each pion if necessary.
We have checked the validity of this prescription by comparing two
different calculations performed by the JAM code.  
The first calculation is done by the JAM+JAM calculation which is 
the same as the AMD+JAM calculation described above but
Eq.~(\ref{eq:fn0}) is solved also by JAM by turning off all the
inelastic $NN$ collisions.  
The result is compared with the standard
JAM calculation which solves all the particles as usual.  With the
present prescription for the conservation laws, the pion
multiplicities and the pion ratios in the two calculations agree well
within the errors of about 10\% and 2\%, respectively, at the incident
energy 300 MeV/nucleon.  These agreements get worse slightly at the
incident energy 400 MeV/nucleon.

The incompleteness of this prescription is the dominant origin of the
violation of the energy conservation in the AMD+JAM calculation, which
can be estimated by the JAM+JAM calculation.  It turned out that the
total energy per baryon is higher than the initial value by about 2 MeV on
average at $t\approx 20$ fm/$c$ in the collisions at 300 MeV/nucleon.
This average increase of the total energy seems roughly consistent
with the above-mentioned 10\%-overestimation of the pion multiplicity.
In fact, the pion multiplicity increases by 13\% in
the standard JAM calculation when the incident energy is raised from
300 to 310 MeV/nucleon, corresponding to the 2.3-MeV increase of the
total energy per baryon.

\subsection{AMD}

AMD \cite{AMD} describes the dynamics of a many-nucleon system by the
time evolution of a Slater determinant of Gaussian wave packets
\begin{equation}
\langle\bm{r}|\varphi_j\rangle = e^{-\nu(\bm{r}-\bm{Z}_j/\sqrt{\nu})^2}\chi_{\alpha_j},
\end{equation}
where the wave packet centroid is denoted by $\bm{Z}_j$ which is a
complex vector, and the spin-isospin state $\chi_{\alpha_j}$ takes
$p\uparrow$, $p\downarrow$, $n\uparrow$ and $n\downarrow$.  The width
parameter is chosen to be $\nu=(2.5\ \mathrm{fm})^{-2}$ as usual.  The
corresponding phase-space distribution is
\begin{multline}
\label{eq:amd-wigner}
f_\alpha(\bm{r},\bm{p})=8\sum_{j\in\alpha}\sum_{k\in\alpha}
e^{-2\nu(\bm{r}-\bm{R}_{jk})^2}
\\\times e^{-(\bm{p}-\bm{P}_{jk})^2/2\hbar^2\nu}B_{jk}B^{-1}_{kj}
\end{multline}
with $\bm{R}_{jk}=(\bm{Z}_j^*+\bm{Z}_k)/\sqrt{\nu}$,
$\bm{P}_{jk}=2i\hbar\sqrt{\nu}(\bm{Z}_j^*-\bm{Z}_k)$ and
$B_{jk}=\langle\varphi_j|\varphi_k\rangle$.

\subsubsection{Mean field term}
The mean field term $\dot{f}_\alpha^{\text{MF}}[f_N,0,0]$ is given by
the equation of motion for the wave packet centroids $\{\bm{Z}_j\}$
derived from the time-dependent variational principle.

In the present calculations, we employ the Skyrme SLy4 force
\cite{SLy4} as the effective interaction with the spin-orbit term
omitted.  The corresponding nuclear-matter incompressibility is
$K=230$ MeV at the saturation density
$\rho_0=0.160\ \mathrm{fm}^{-3}$.  The nuclear-matter symmetry energy
at $\rho_0$ is $S_0=32.0$ MeV with the slope parameter $L=46$ MeV
(called `asy-soft' or soft symmetry energy in Sec.~\ref{sec:results}).
In order to study the effect of the density dependence of the symmetry
energy, we also perform calculations with a force obtained by changing
the density dependent term in the SLy4 force
\begin{equation}
v_\rho^{(L=46)}=\tfrac{1}{6}t_3(1+x_3P_\sigma)
\rho(\bm{r}_1)^\alpha\delta(\bm{r}_1-\bm{r}_2)
\end{equation}
to
\begin{multline}
v_\rho^{(L=108)}=\tfrac{1}{6}t_3(1+x_3'P_\sigma)
\delta(\bm{r}_1-\bm{r}_2)\rho(\bm{r}_1)^\alpha
\\
+\tfrac{1}{6}t_3(x_3-x_3')\rho_0^\alpha P_\sigma\delta(\bm{r}_1-\bm{r}_2).
\end{multline}
By choosing $x_3'=-0.5$, we have a force corresponding to $L=108$ MeV
(callded `asy-stiff' or stiff symmetry energy) with the same equation
of state of symmetric nuclear matter and with the same $S_0$ as the
original SLy4 force.

The Skyrme-type prarametrization of the effective interaction is
advantageous for the efficient AMD computation.  However, the Skyrme
forces have a quadratic momentum dependence of the mean field which is
not valid at high energy collisions at several hundred MeV/nucleon.
Therefore, the momentum dependence is corrected for the present
calculations in a similar way to Ref.~\cite{GBDG} in BUU-type
calculations.  The detailed formulation in the case of AMD is given in
the Appendix \ref{sec:mdcorr}.

\subsubsection{Collision term with and without clusters}
The two-nucleon collision process corresponds to the collision term
$I_N^{(\lambda=0)}[f_N,0,0]$.  In AMD, a two-nucleon collision is
treated as a stochastic transition from an AMD state $|\Phi_i\rangle$
to another AMD state $|\Phi_f\rangle$ specified by the relative
momentum between the scattered two nucleons
$(p_{\text{rel}},\Omega)$. The transition rate is expressed as
\begin{equation}
vd\sigma=\frac{2\pi}{\hbar}|\langle\Phi_f|V|\Phi_i\rangle|^2\delta(E_f-E_i)
\frac{p_{\text{rel}}^2dp_{\text{rel}}d\Omega}{(2\pi\hbar)^3}.
\label{eq:transitionrate}
\end{equation}
In general, medium modification is introduced for the scattering
matrix elements.  However, in the present calculations, we employ the
matrix elements in the free space.  It should be noted that some
medium effect still exists in the $p_{\text{rel}}$-dependence of the
final state energy $E_f$ in a similar way to a BUU calculation of
Ref.\ \cite{BUU}.  The Pauli blocking for the scattered nucleons is
taken into account.

In the usual treatment of two-nucleon collisions, only the states of
the scattered two nucleons are changed in the final state
$|\Phi_f\rangle$ (see Ref.~\cite{AMD} for the precise description of
the method which employs `physical coordinates').  On the other hand,
an extension has been introduced \cite{Ono:NN2012} to allow direct
formation of light clusters with $A=2,3$ and 4 in the final state
$|\Phi_f\rangle$.  Namely, in the calculation with cluster
correlations, when two nucleons $N_1$ and $N_2$ collide, we consider
the process
\begin{equation}
N_1+N_2+B_1+B_2\rightarrow C_1+C_2
\end{equation}
in which each of the scattered nucleons $N_j$ ($j=1,2$) may form a
cluster $C_j$ with a spectator particle $B_j$.  This process includes
the collisions without cluster formation as the special case of
$C_j=N_j$ with empty $B_j$.  The transition rate of the
cluster-forming process is given by Eq.~(\ref{eq:transitionrate}) with
the suitable choice of the final state $|\Phi_f\rangle$.  When a
cluster is formed, the corresponding wave packets are placed at the
same phase-space point, i.e., the cluster internal state is
represented by the harmonic-oscillator $(0s)^n$ configuration.
Denoting the initial and final states of the $N_j+B_j$ system by
$|\varphi_j\rangle$ and $|\varphi_j'\rangle$, respectively, we have
the transition rate
\begin{multline}
vd\sigma = \frac{2\pi}{\hbar}
|\langle\varphi_1'|\varphi_1^{\bm{q}}\rangle|^2
|\langle\varphi_2'|\varphi_2^{-\bm{q}}\rangle|^2
\\
\times|M|^2\delta(E_f-E_i)
\frac{p_{\text{rel}}^2dp_{\text{rel}}d\Omega}{(2\pi\hbar)^3},
\label{eq:transitionrate-cluster}
\end{multline}
where
$|\varphi_j^{\pm\bm{q}}\rangle = e^{\pm
  i\bm{q}\cdot\bm{r}_j}|\varphi_j\rangle$
are the states after the momentum transfer $\pm\bm{q}$ to the nucleons
$N_j$ ($j=1,2$), and $(p_{\text{rel}},\Omega)$ is the relative
momentum between $N_1$ and $N_2$ in these states.  The matrix element
$|M|^2$ is the same as for the usual two-nucleon collisions.  We use
an average value of $|M|^2$ evaluated at $p_{\text{rel}}$ and that
evaluated at the initial relative momentum.

The actual situation of a two-nucleon collision requires more
considerations because there are many possible ways of forming a
cluster for each $N$ of the scattered nucleons $N_1$ and $N_2$.  For a
scattered nucleon $N$, we first consider the possibility that $N$ may
form a cluster with one of the nucleons $\{B_k;\ k=1,2,\dots\}$ which
have the same spin-isospin state.  This spin-isospin state that is
studied first is randomly decided.  The cluster-formed state is
denoted by $|\Phi_k'\rangle$ which is obtained, by first changing the
state to $|\Phi^{\bm{q}}\rangle$ by the momentum transfer $\bm{q}$ to
$N$, and then moving the two wave packets of $N$ and $B_k$ to the same
phase-space point without changing their center of mass.  Since the
different final states are not orthogonal
$\mathcal{N}_{kl}=\langle\Phi_k'|\Phi_l'\rangle\ne\delta_{kl}$, the
probability that $N$ forms a cluster with one of $\{B_k\}$ should be
calculated as
\begin{gather}
P=\sum_{kl}\langle\Phi^{\bm{q}}|\Phi_k'\rangle
\mathcal{N}^{-1}_{kl}
\langle\Phi_l'|\Phi^{\bm{q}}\rangle
=\sum_{k}|v_k|^2,
\\
v_k=\sum_{l}\mathcal{N}^{-1/2}_{kl}\langle\Phi_l'|\Phi^{\bm{q}}\rangle,
\end{gather}
This probability is calculated with an approximation that the
many-body state is a direct product of wave packets centered at the
`physical coordinates' \cite{AMD}.  With the calculated probability
$P$, a cluster will be formed with one of $\{B_k\}$.  It is somewhat
arbitrary which one of $\{B_k\}$ should be chosen with what probability.
In the present calculation, we choose $B_k$ with the relative weight
$|v_k|^{2\gamma}$ with the parameter $\gamma=2.0$.  With the rest of
the probability ($1-P$), $N$ does not form a cluster with a nucleon of
this spin-isospin state.  The procedure is repeated for other
spin-isospin states for $\{B_k\}$.  The particle $N$ should be
regarded as a cluster, instead of a scattered nucleon, if a
(sub)cluster has been already formed in previous steps of the
repetition.  Thus the formation of light clusters is considered up to
an $\alpha$ particle.  This procedure determines the probability
$P_{C_1C_2}(p_{\text{rel}},\Omega)$ for the combination of final
clusters $(C_1,C_2)$ as a function of the momentum transfer $\bm{q}$
or $(p_{\text{rel}},\Omega)$.  It satisfies the normalization
$\sum_{C_1C_2}P_{C_1C_2}(p_{\text{rel}},\Omega)=1$. The factor
$|\langle\varphi_1'|\varphi_1^{\bm{q}}\rangle|^2|\langle\varphi_2'|\varphi_2^{-\bm{q}}\rangle|^2$
in Eq.~(\ref{eq:transitionrate-cluster}) should be replaced by
$P_{C_1C_2}(p_{\text{rel}},\Omega)$.

Even when the cluster formation is introduced, the many-body state is
always represented by an AMD wave function which is a Slater
determinant of nucleon wave packets.  The time evolution of the
many-body state is solved just as usual without depending on whether
some of the wave packets form clusters due to collisions in the past
(except for the cluster-cluster binding process in the next
paragraph).  This is in contrast to BUU by Danielewicz \textit{et
  al.}~\cite{Danielewicz:1991} where clusters are treated as new
particle species.  In our approach, a nucleon in a formed cluster may
collide with some other nucleon so that the cluster is broken.  It may
be the case that the scattered nucleon forms the same cluster as
before, so that an elastic scattering of the cluster is possible.  All
of these kinds of processes are based on the nucleon-nucleon
scattering matrix elements, without introducing parameters to control
individual channels of cluster formation.  In the present calculations
with cluster correlations, however, the overall cluster production
probability is suppressed, when the momentum transfer is extremely
small, by a factor $1-\exp[-\bm{q}^2/(50\ \text{MeV}/c)^2]$.

It has been found that we should take account of
the correlations to form heavier fragments
via coalescence of light clusters
on top of the usual time evolution of AMD \cite{Ono:NN2012}.
This option of improvement has
been turned on in the present calculations even though it will not
strongly influence the following discussions in high energy
collisions.  The details are described in Appendix \ref{sec:coacc} for
completeness.

It is experimentally clearly known that the clusters are important in
many situations of heavy-ion collisions even though the incident
energy is relatively high.  For example, the FOPI data \cite{FOPI}
show that only 21\% of the total protons in the Au + Au system are
emitted as free protons in central collisions at 250 MeV/nucleon, and
all the other protons are bound in light clusters and heavier
fragments.  AMD without cluster correlations overestimates the proton
multiplicity as many other transport models do.  On the other hand,
the AMD calculation with cluster correlations can reproduce this
feature very well as shown in Ref.~\cite{Ono:NN2012} for systems
including the Au + Au central collisions at several hundred
MeV/nucleon.  Therefore, the calculation with clusters is believed to
be much closer to the realistic case than without clusters.  In the
following section, however, we are going to show the results of both
calculations with and without clusters, which are useful for the
purpose to study the dependence of the pion production on the nucleon
dynamics.

\subsubsection{Test particles}

The information of nucleon dynamics calculated by AMD is sent to the
JAM calculation in the form of a set of test particles
$(\bm{r}_1,\bm{p}_1)$,
$(\bm{r}_2,\bm{p}_2)$,\dots,$(\bm{r}_A,\bm{p}_A)$.  One test particle
per nucleon is generated following the distribution function defined
by Eq.~(\ref{eq:amd-wigner}).  It should be noted that all kinds of
quantum effects from the antisymmetrization of the many-body state are
contained in this distribution function (or the Wigner function).  In
particular, it is not positive definite, and therefore in the
phase-space region of $f_\alpha(\bm{r},\bm{p})<0$ the probability has
to be replaced by zero, which can potentially introduce some
inaccuracy of the test-particle representation.  To check the
accuracy, we compared the density profile for the ground state of the
Au nucleus, to find no visible difference between the distribution of
the generated test particles and the exact density profile.
Therefore, this method of test particles should be sufficiently
accurate in highly excited situations during heavy-ion collisions.
The method to generate the test particles is described in
Appendix~\ref{AppC}.

The set of test particles is sent to the JAM calculation at every 2
fm/$c$.  We have checked that the result does not change when it is
sent at every 1 fm/$c$.

The AMD calculation is much more time consuming than other transport
models such as JAM.  We typically generated 1000 AMD events for each
case of the present calculations.  However, we improve the statistics
for the pion production by generating 500 JAM events from the same AMD
event.  As we will see in the next section, a sufficient statistical
accuracy is obtained for the pion production with this limited number
of AMD events.

\subsection{JAM}
JAM is a transport model which is developed by Nara \textit{et
  al.}~\cite{JAM}. This model has been successfully applied to
high-energy collisions up to more than one hundred GeV/nucleon.  In
this model, in the energy domain relevant for the present work, the
hadron-hadron reactions are treated by the cross sections based on
experimental data and the detailed balance.
In particular, in the present work, isospin symmetry is assumed.  The
cross section for the $NN\rightarrow N\Delta$ process at the c.m.\
energy $\sqrt{s}$ is written as~\cite{Teis:1996kx,Weil:2012ji}
\begin{multline}
\sigma_{NN \rightarrow N\Delta} = \frac{C_I}{p_{i} s} 
\frac{|\mathcal{M}|^2}{16\pi}\\
\times
\int dm \frac{2}{\pi} \frac{m^2 \Gamma(m)}{(m^2 -m^2_{\Delta})^2+ m^2
\Gamma(m)^2 } p_f(m)
\label{eq_sigma}
\end{multline}
where $p_i$ and $p_f(m)$ are the initial and final momenta in the
c.m.\ frame, and the Clebsh-Gordan factor is $C_I=\frac{1}{4}$ or
$\frac{3}{4}$.
The matrix element $|\mathcal{M}|$ is assumed as 
\begin{equation}
|\mathcal{M}|^2=A\frac{s\Gamma_\Delta^2}{(s-m_\Delta^2)^2+s\Gamma_\Delta^2}
\end{equation}
with $\Gamma_\Delta=0.118$ GeV and $m_\Delta=1.232$ GeV.  
This is a
similar parametrization to UrQMD in Ref.~\cite{UrQMD}.  The constant
$A/16 \pi=64400$ mb GeV$^2$
is determined to fit the data of the $NN\rightarrow NN\pi$ cross
sections.  
The distribution of the mass $m$ of the $\Delta$ resonance is
determined by the integrand of Eq.~(\ref{eq_sigma}).
$\Gamma(m)$ is the decay width for $\Delta\rightarrow N\pi$
parametrized as in Refs.~\cite{JAM,UrQMD}.  In order to fit the data
precisely near the pion threshold, non-resonant contributions and/or
some components that violates isospin symmetry are necessary for the
$NN\rightarrow NN\pi$ processes, which are ignored, however, in the
present work so that the pions are produced only through the formation
of resonances.  In our JAM calculation, the mean field for nucleons is
not included.

\begin{figure}
\begin{center}
\includegraphics[width=8.0cm, height=6.8cm]{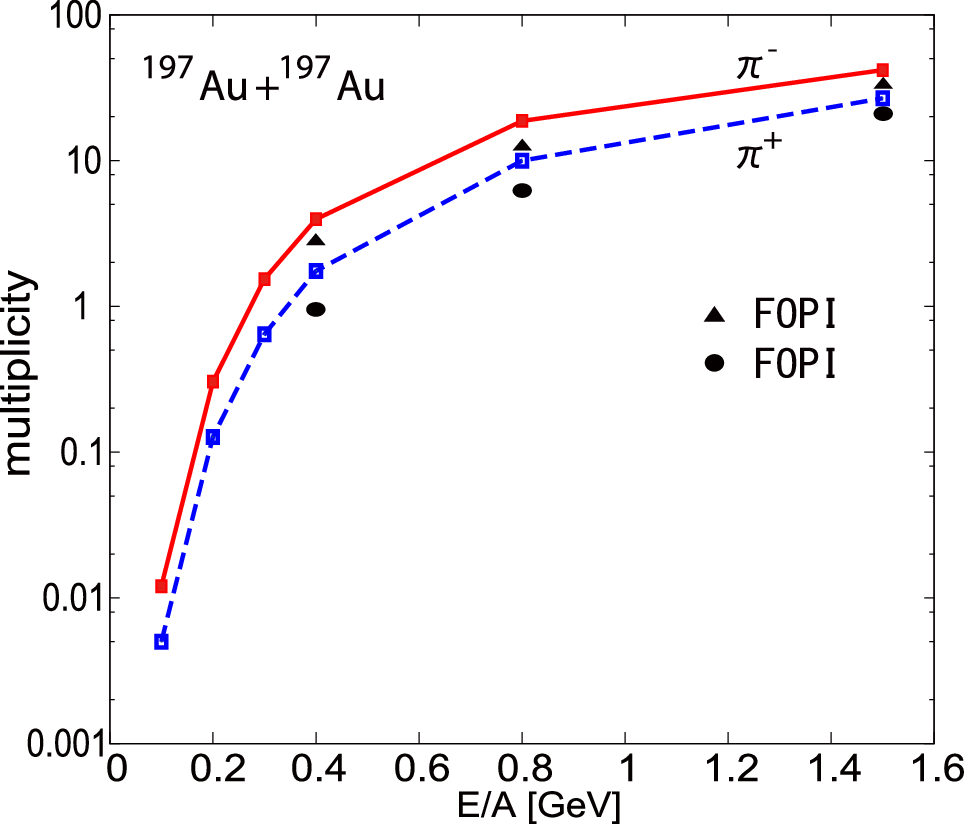}
\caption{ Incident energy dependence of the charged pion
  multiplicities in central Au + Au collisions.  The two lines
  indicate the multiplicities of $\pi^-$ and $\pi^+$, respectively,
  calculated by JAM.  Symbols represent the experimental data of the
  FOPI Collaboration~\cite{FOPI}.}
\label{Edep_JAM}
\end{center}
\end{figure}

\begin{figure}
\begin{center}
\includegraphics[width=6.5cm,height=7.5cm]{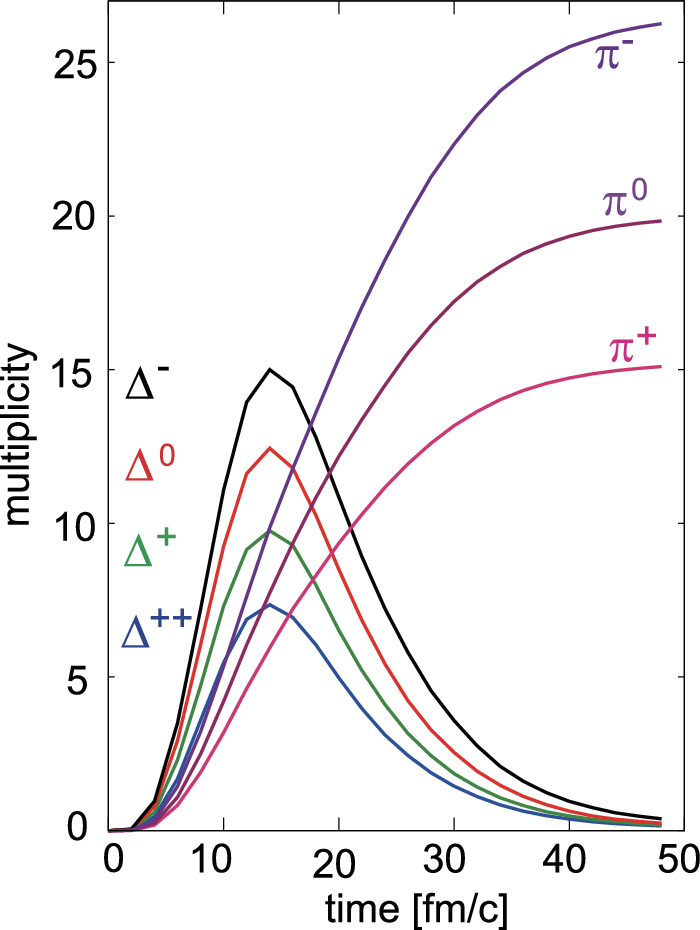}
\caption{Time evolution of the numbers of pions and $\Delta$
  resonances calculated by JAM for the central Au + Au collisions at 1
  GeV/nucleon with the impact parameter $b=0$.}
\label{Tdep_JAM}
\end{center}
\end{figure}

To know how the JAM calculation reproduces the pion multiplicity of
the experimental data, we have calculated Au + Au collisions at
various incident energies for the impact parameters $0<b<2$ fm.
Figure~\ref{Edep_JAM} shows the energy dependence of the pion
multiplicity.  The solid lines indicate the pion multiplicities
calculated by JAM.  The points show the experimental data taken from
the FOPI measurement of Au + Au collisions at 0.4, 0.8 and 1.5
GeV/nucleon \cite{FOPI}.  We find that the JAM calculation almost
reproduces the experimental data of pion multiplicities reasonably
well.  However, the calculation overestimates the pion multiplicities
in particular at lower energies.  This is probably because the JAM
calculation here does not include the mean field potential, and
therefore more energy is available for particle production without the
cost of energy to compress the
system~\cite{Danielewicz:1991,Kruse:1985hy}.

We have also checked the the time evolution of the numbers of pions
and $\Delta$ resonances in the JAM calculations as shown in
Fig.~\ref{Tdep_JAM} for the central Au + Au collisions at 1
GeV/nucleon.  The result shows a similar behavior to a relativistic
mean-field transport calculation reported in Fig.~1 of
Ref.~\cite{Ferini}.

Thus JAM calculation provides a reasonable description for the pions
and $\Delta$ resonances produced in collisions at around 1
GeV/nucleon.  It should be noted that in the AMD+JAM calculation the
mean field for nucleons is taken into account in AMD and the ignored
mean field in JAM does not actually influence the results.

The production and absorption reactions for $\Delta$ and pions occur
in the JAM calculation as in the free space without medium
modification for the thresholds, while nucleons feel potential in the
AMD calculation.  This corresponds to assuming that the potentials
$U_\tau^{(\Delta)}$ and $U_\tau^{(\pi)}$ for $\Delta$ and pions are
related to the isospin-dependent nucleon potential
$U_\tau^{(N)}=U_0+\tau U_{\text{sym}}$ as
\begin{equation}
  U_\tau^{(\Delta)}=U_0 + \tau U_{\text{sym}},\quad
  U_\tau^{(\pi)}=\tau U_{\text{sym}},
\end{equation}
where $\tau$ is the isospin component.  This is equivalent to the
choice in the pBUU calculation of Ref.~\cite{pBUU} if the momentum
dependence is ignored.  It should be noted that other transport
calculations take a different choice \cite{Li,Gaitanos:2004}.  The
details of the in-medium effects for $\Delta$ and pions may influence
the pion yields as investigated in equilibrium calculations
\cite{JunXu:2010,JunXu:2013}.  On the other hand, recently, the
effects of the completely unknown symmetry (isovector) potential of
the $\Delta$ resonance on the pions in heavy-ion collisions have been
studied in Ref \cite{Li:2015hfa}.  It has been reported that these
effects are negligible except for at deeply subthreshold energies and
thus the pion ratio is still a good probe to investigate the
high-density symmetry energy.

\section{\label{sec:results}Results}
We calculated collisions of ${}^{132}\mathrm{Sn}+{}^{124}\mathrm{Sn}$
at 300 MeV/nucleon for the impact parameters $0<b<1$ fm.  At the
initial time $t=0$, the centers of the two nuclei are separated by 15
fm.  In order to investigate the relation between the high-density
symmetry energy, the nucleon dynamics in a compressed neutron-rich
system and the emitted pions, we are going to compare the results from
five different cases as follows,
\begin{enumerate}
\item AMD + JAM with clusters (asy-soft)
\item AMD + JAM with clusters (asy-stiff)
\item AMD + JAM without clusters (asy-soft)
\item AMD + JAM without clusters (asy-stiff)
\item JAM (no mean field).
\end{enumerate}
The first two cases are calculated with AMD with cluster correlations,
and the next two cases are calculated without cluster correlations.
For each of them, we have calculated with two different effective
interactions (`asy-soft' and `asy-stiff') for different density
dependence of symmetry energy.  We have also performed a standard JAM
calculation without combining with AMD in order to clarify the effect
of the mean field and the symmetry energy by the comparison with the
AMD+JAM calculations.

\subsection{Neutron-proton dynamics}
\begin{figure*}
\begin{center}
 \includegraphics[width=17.2cm,height=12cm]{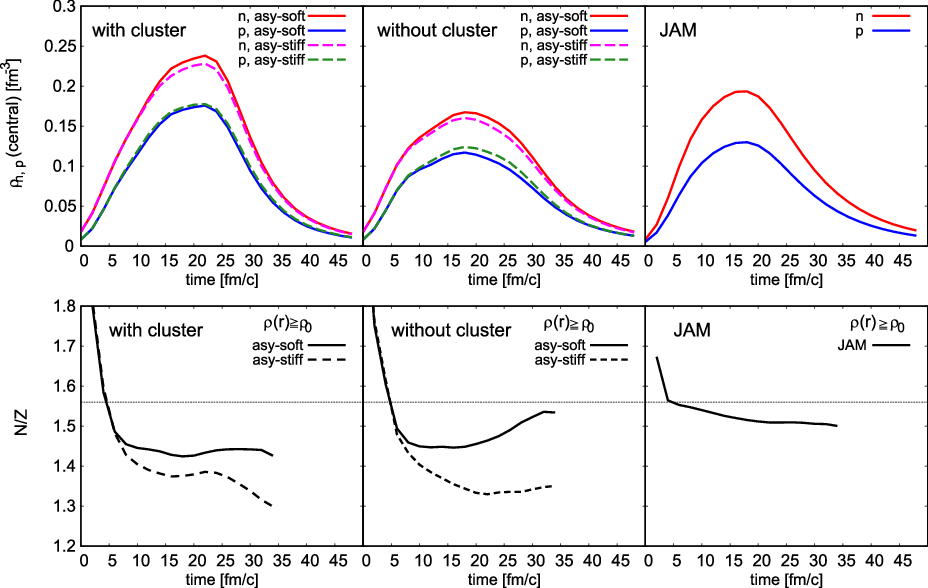}
\caption{Time evolution of the neutron and proton densities in the
  central part of the system within the radius of 2 fm (upper) and the
  ratio of neutrons and protons (lower) in central collisions of
  ${}^{132}\mathrm{Sn}+{}^{124}\mathrm{Sn}$ at 300 MeV/nucleon.
The left and middle
    panels show the results of AMD+JAM with clusters and without
    clusters, respectively. The right panels are for the simple JAM
    calculation.  Two different lines (solid and dotted) correspond
  to the two different density dependence of symmetry energy.  The
  horizontal lines in the bottom panels represent the ratio of the
  total system $(N/Z)_{\rm sys}=1.56$. }
\label{dens}
\end{center}
\end{figure*}

To see the dynamics of neutrons and protons in the five different
calculations, Fig.~\ref{dens} shows some information on the time
evolution of the densities of protons and neutrons. The upper panels
show the neutron and proton densities in the radius of 2 fm from the
center-of-mass of the system.  The maximum density
$\rho=\rho_n+\rho_p\sim 2\rho_0$ is reached at $t \simeq 20$~fm/$c$.
However, we find that the maximum density is higher with cluster
correlation than without it.  The time of the maximum density also
depends on the cluster correlation.  One of the possible reasons for
this is that the cluster formation has an effect to gather nucleons
spatially so that the compression of the central part of the system
continues longer.  In the case of JAM calculation, the higher maximum
density is reached at an earlier time than in the AMD without
clusters, which is reasonable because the mean-field potential for
nucleons is not included in JAM.

The lower panels of Fig.~\ref{dens} show the time evolution of the
neutron-to-proton ratio $N/Z$ calculated for the central region within
a radius $r_0$ from the center of mass.  The radius $r_0$ is
determined at each time in each event by the condition
$\rho(r_0)\approx\rho_0$, where the spherically averaged density
$\rho(r)$ is evaluated by using the set of test particles.  The
results with the soft and stiff symmetry energies are shown by the
solid and dashed lines, respectively, for each case of with and
without cluster correlation.  In all the cases, the $N/Z$ ratio of the
compressed part (at $t\gtrsim10$ fm/$c$) becomes smaller than the that
of the total system $(N/Z)_{\text{sys}}=1.56$, which is consistent
with the symmetry energy effect that does not favor high-density
neutron-rich matter.  We can see clearly that this effect to reduce
$N/Z$ of the compressed part is stronger with the stiff symmetry
energy.  This symmetry energy effect is consistent with the results of
other transport models \cite{Li} at least qualitatively.  However,
there may be model dependence in the quantitative values of $N/Z$.  In
fact, in our calculations here, the symmetry energy effect is stronger
without cluster correlation than with cluster correlation.

\subsection{$\Delta$ and pions}
\begin{figure}
\begin{center}
\includegraphics[width=8.0cm, height=15.0cm]{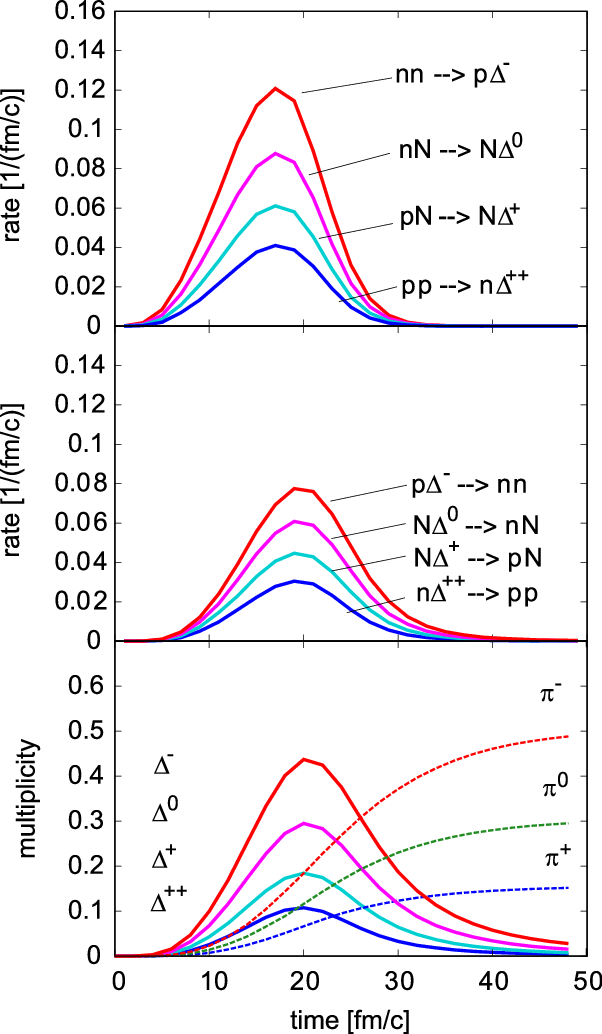}
\caption{ Reaction rates of $\Delta$ production (upper) and $\Delta$
  absorption (middle), and the numbers of existing $\Delta$ resonances
  and pions (lower) as functions of time, in the AMD+JAM calculation
  with clusters (asy-soft) for central collisions of
  ${}^{132}\mathrm{Sn}+{}^{124}\mathrm{Sn}$ at 300 MeV/nucleon. 
For the production and absorption of $\Delta^+$ and $\Delta^0$, the
line shows the sum of the reaction rates with a neutron ($N=n$) and
with a proton ($N=p$). 
}
\label{delta}
\end{center}
\end{figure}

\begin{figure}
\begin{center}
\includegraphics[width=8.5cm, height=6.8cm]{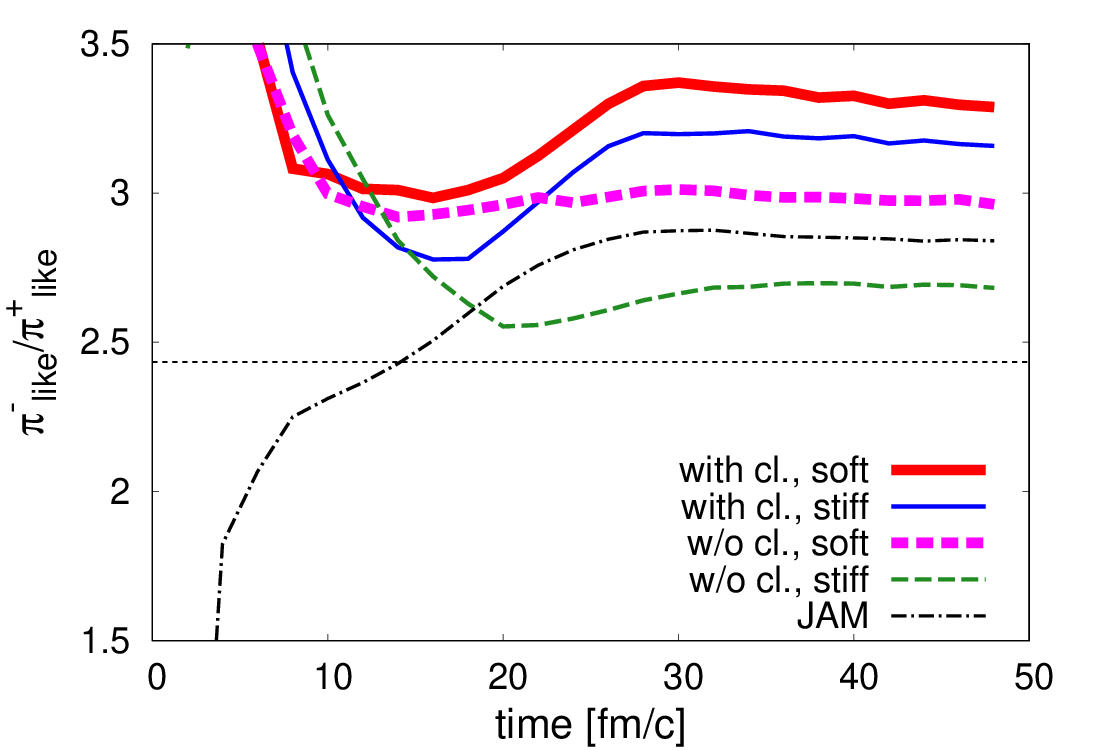}
\caption{The time evolution of the
  $\pi^-_{\text{like}}/\pi^+_{\text{like}}$ ratios in central
  collisions of ${}^{132}\mathrm{Sn}+{}^{124}\mathrm{Sn}$ at 300
  MeV/nucleon in the five cases of calculation.  The horizontal line
  represents the $(N/Z)^2_{\rm sys}$ ratio of the total
  system. 
}
\label{pion-like_ratio}
\end{center}
\end{figure}

In Fig.~\ref{delta}, we show the reaction rates of $\Delta$ production
($NN\rightarrow N\Delta$) and absorption ($N\Delta\rightarrow NN$) in
the upper and middle panels, respectively.  The time evolution of these
rates for $\Delta^{-}$, $\Delta^{0}$, $\Delta^{+}$ and $\Delta^{++}$
are shown by the four lines.  We can see that about 70\% of the
produced $\Delta$ resonances are absorbed and turned back to nucleons.
We also show the numbers of existing $\Delta$ resonances and pions 
in the lower panel of Fig.~\ref{delta}.  

In Fig.~\ref{pion-like_ratio}, we show the time evolution of the
$\pi^{-}_{\rm like}/\pi^{+}_{\rm like}$ ratio for the five different
cases.  
The $\pi_{\rm like}$ particles are defined,
including the $\Delta$ resonances depending on the branching ratio to
decay into the pion, as
\begin{align}
 \pi^{-}_{\rm like} &= \pi^{-} + \Delta^{-} + \tfrac{1}{3}\Delta^{0},\\
 \pi^{0}_{\rm like} &= \pi^{0} + \tfrac{2}{3}\Delta^{0} + \tfrac{2}{3}\Delta^{+},\\
 \pi^{+}_{\rm like} &= \pi^{+} + \Delta^{++} + \tfrac{1}{3}\Delta^{+}.
\end{align}
Our calculation predicts that the evolution of the pion-like
ratio has a dependence on the symmetry energy.  The pion-like ratio
calculated with the soft symmetry energy is larger than that with the
stiff symmetry energy in both of calculations with and without cluster
correlations.  This result seems to be similar to the predictions
reported in Refs.~\cite{Li,IBUU04} qualitatively.  We also find that
the pion-like ratio depends on the cluster correlations and that the
symmetry-energy effect appears in the pion-like ratio stronger in the
case without cluster correlations.  This is consistent with the result
of the neutron-proton dynamics as shown in the lower panels of
Fig.~\ref{dens}, suggesting a possibility that the pion production is
really related to the neutron-proton dynamics.

The pion-like ratio reaches the final $\pi^-/\pi^+$ value at around
$t \simeq 30$ fm/$c$.  In all the five cases, the predicted final
$\pi^-/\pi^+$ ratios become larger than $(N/Z)_{\text{sys}}^2=2.4336$
of the total system, which is consistent with the experimental
observation for Au + Au system at 400 MeV/nucleon \cite{Reisdorf:2007}
and is suggesting that the relation $\pi^-/\pi^+\approx(N/Z)^2$ does
not hold if $(N/Z)$ is taken from Fig.~\ref{dens}.  On the other hand,
the behaviors of the pion-like ratio before $t \simeq 30$ fm/$c$ are
complicated.  The origin of these behaviors will be better understood
through the analysis in the next subsections.

\subsection{Relation of nucleon dynamics and $\Delta$ production}

\begin{figure}
\begin{center}
\includegraphics[width=8.5cm, height=6.8cm]{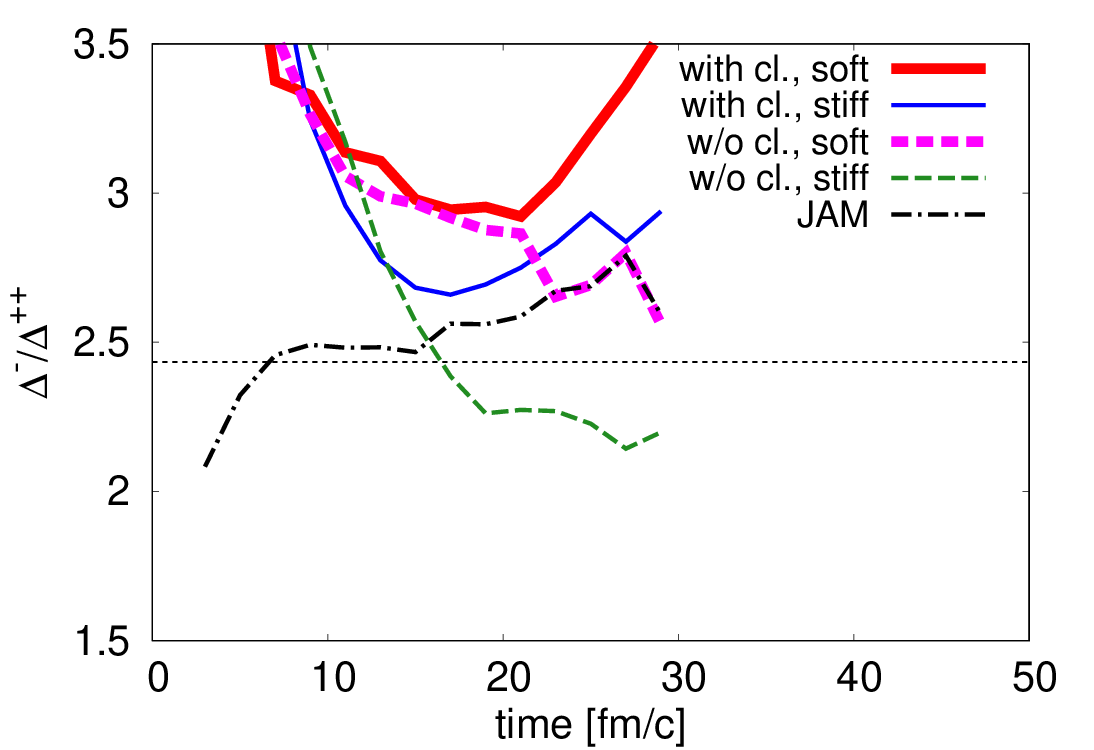}
\caption{The time evolution of the $\Delta^{-}/\Delta^{++}$ ratio of
  the $\Delta$ production rates.  The five different lines show the
  calculations.  The horizontal line represents the
  $(N/Z)^2_{\rm sys}$ ratio of the total system. }
\label{fig:delta-production-ratio}
\end{center}
\end{figure}

\begin{figure*}
\begin{center}
\includegraphics[width=8.6cm, height=6.5cm]{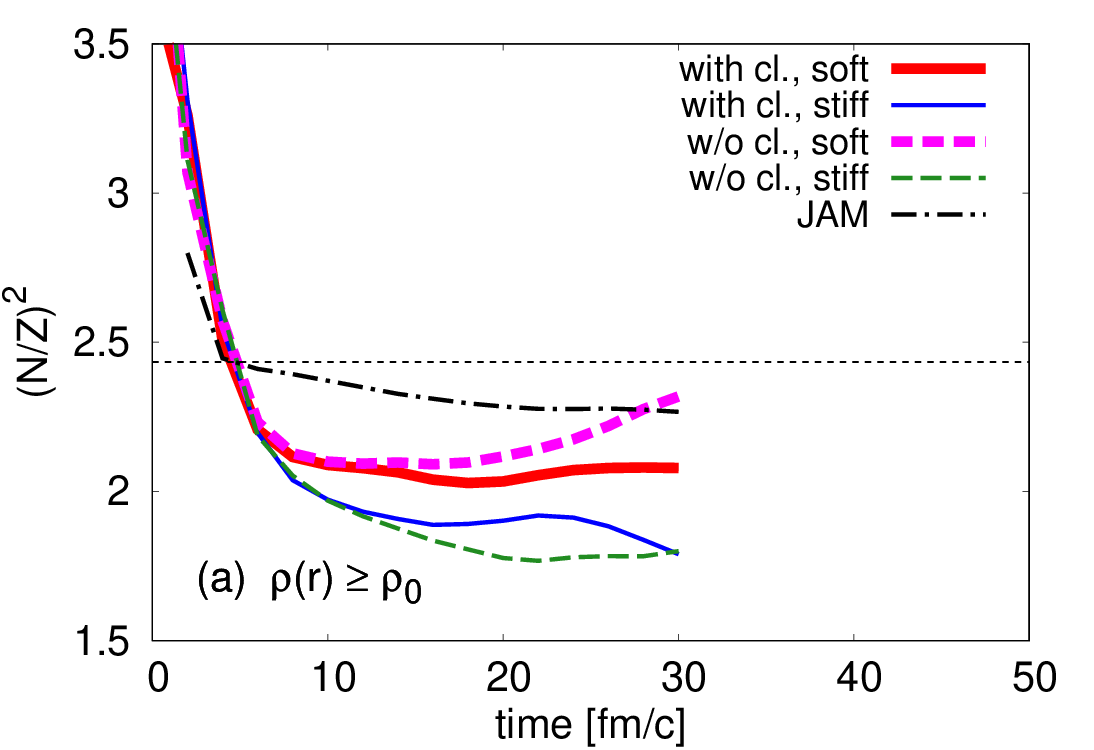}
\includegraphics[width=8.6cm, height=6.5cm]{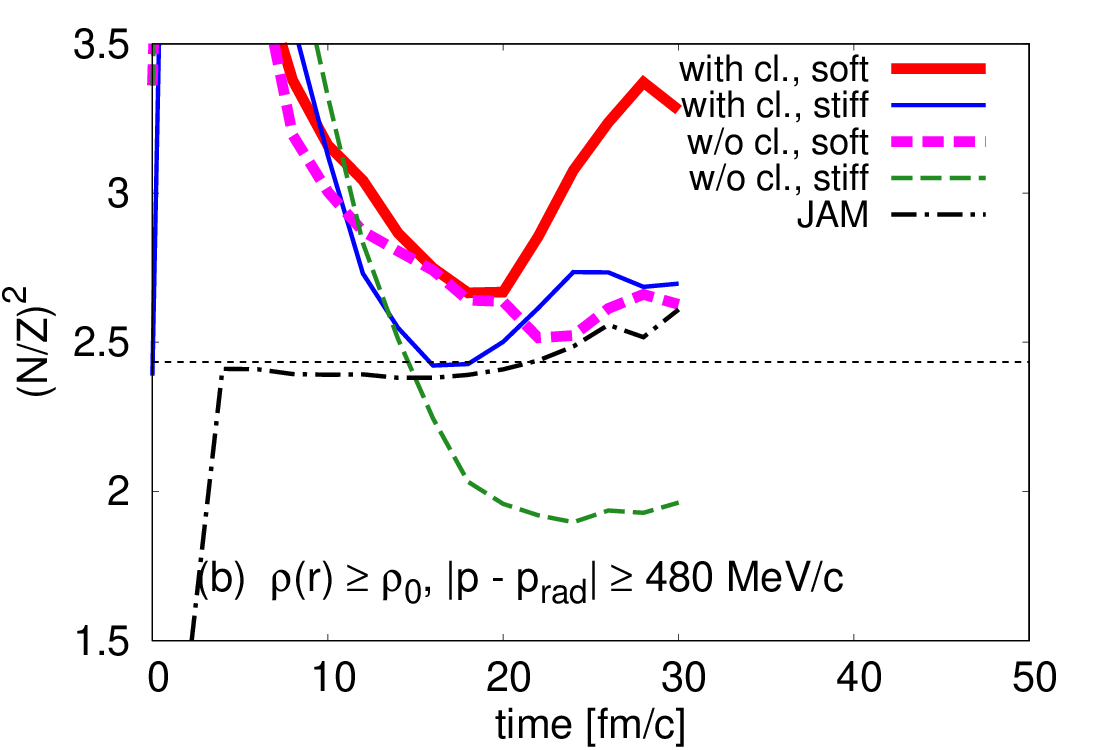}
\caption{ The time evolution of the squared ratio of neutron and
  proton $(N/Z)^2$.  The left and right figures show the $(N/Z)^2$
  ratio calculated for the nucleons in the condition (a) and (b),
  respectively.  See details in the text. 
}
\label{fig:np-ratio-squared}
\end{center}
\end{figure*}

As mentioned above, 
the calculated pion ratio becomes larger than
the $(N/Z)^2$ ratio of the compressed part of the system.
To find the origin of this effect, we investigate
what kind of information of neutrons and protons
is carried by $\Delta$ resonances.
Since $\Delta^{-}$ and $\Delta^{++}$ are produced only by the
$nn\rightarrow p\Delta^{-}$ and $pp\rightarrow n\Delta^{++}$
reactions, respectively, we expect that the $\Delta^{-}/\Delta^{++}$
ratio of the production rates of these resonances should be most
directly linked to some kind of $(N/Z)^2$ ratio of nucleons.  Figure
\ref{fig:delta-production-ratio} shows the ratio of the production
reaction rates of $\Delta^{-}$ and $\Delta^{++}$ as functions of time
for the five different calculations.  The $\Delta$ production is
peaked around $t=15$-20 fm/$c$ as shown in Fig.~\ref{delta}, and the
ratio does not have a significant meaning at very early and late
times.  This $\Delta$ production ratio is compared with
Fig.~\ref{fig:np-ratio-squared}(a) which shows the same information as
in the lower panels of Fig.~\ref{dens} but shows the squared ratio
$(N/Z)^2$.  We find clearly that the $\Delta$ production ratio is much
larger than this $(N/Z)^2$ ratio in the high-density region of the
system, except for the JAM case in which both ratios are close to the
$(N/Z)^2_{\text{sys}}$ ratio of the total system.  The relative
ordering of the ratios for the five cases also disagrees between
Fig.~\ref{fig:delta-production-ratio} and
Fig.~\ref{fig:np-ratio-squared}(a).  This result shows that the
$\Delta$ production, and therefore the pion production, are not simply
linked to the $N/Z$ ratio of the high-density part of the system.

Each panel of Fig.~\ref{fig:np-ratio-squared} shows
the time evolution of the
squared neutron-to-proton ratio $(N/Z)^2$
calculated for the nucleons that satisfy the following
condition.
\begin{itemize}
\item[(a)] Nucleons in the sphere $\rho(r) \geq \rho_0$ centered
  at the center-of-mass of the system.
\item[(b)] Nucleons with high momenta
  $|\bm{p} - \bm{p}_{\rm rad}| \geq$ $p_{\rm cut}$ in the sphere
  $\rho(r) \geq \rho_0$ centered at the center-of-mass of the system.
  We take $p_{\rm cut}=480$ MeV/$c$.
  The collective radial momentum $\bm{p}_{\rm rad}$ is subtracted from
  the nucleon momentum $\bm{p}$.
\end{itemize}
For the condition (a), we choose the nucleons in the high-density
central region within a radius $r_0$ from the center of mass.  The
radius $r_0$ is determined by the condition $\rho(r_0)\approx\rho_0$
as described in the subsection~A.  For the condition (b), we choose
only the nucleons with momenta $|\bm{p}-\bm{p}_{\text{rad}}|$ larger
than $p_{\rm cut}=480$ MeV/$c$ in the center-of-mass system, in addition to the
condition (a).  The collective radial momentum
$\bm{p}_{\text{rad}}=p_{\text{rad}}(r)\bm{r}/r$ is subtracted for this
condition, where $p_{\text{rad}}(r)$ is the radial momentum component
averaged for the nucleons on the sphere of the radius $r$.  
It is natural to consider this kind of momentum condition since a sufficient
energy is required to excite a $\Delta$ resonance in a two-nucleon
collision.
Our choice of $p_{\rm cut}$ corresponds to $p_{\rm cut}^2 /m_{N}=245$ MeV
while $m_{\Delta} - m_{N}=293$ MeV.
We have also checked that the changes of $p_{\rm cut}$ by $\pm 20$ MeV/$c$
result in $\pm 3 \%$ differences in $(N/Z)^2$
(for the average values shown in Fig.~\ref{ratio_summary}). 

The panels of Fig.~\ref{fig:np-ratio-squared} correspond to
the results of the time evolution of the
$(N/Z)^2$ ratio calculated for the nucleons
satisfying the conditions (a) and (b), respectively.
We find that when the high momentum condition is imposed,
the $(N/Z)^2$ ratio changes drastically compared to that without the
momentum condition.  The $(N/Z)^2$ ratio in condition (b) becomes larger
than that in condition (a).

By comparing the $\Delta^{-}/\Delta^{++}$ production ratio in
Fig.~\ref{fig:delta-production-ratio} and the $(N/Z)^2$ ratio in
Fig.~\ref{fig:np-ratio-squared}(a) for each condition, we have already
seen that $\Delta^{-}/\Delta^{++}$ and $(N/Z)^2$ do not agree if the
nucleons are selected only by the high-density condition (a).  On the
other hand, in the result of the condition (b),
$\Delta^{-}/\Delta^{++}$ is quite similar to $(N/Z)^2$, i.e., the
relation $\Delta^{-}/\Delta^{++} \simeq (N/Z)^2$ holds as a function
of time.  Thus, we can conclude that $\Delta$ resonances, and
hopefully pions, carry direct information on nucleons in high density
and high momentum region of the one-body phase space.  We also mention
that the agreement is not as perfect as in the case (b) at $t=20$-30
fm/$c$ if the collective radial momentum is not subtracted to define
the condition.

\subsection{From nucleons to pion ratios}
\begin{figure}
\begin{center}
\includegraphics[width=8.6cm, height=7.5cm]{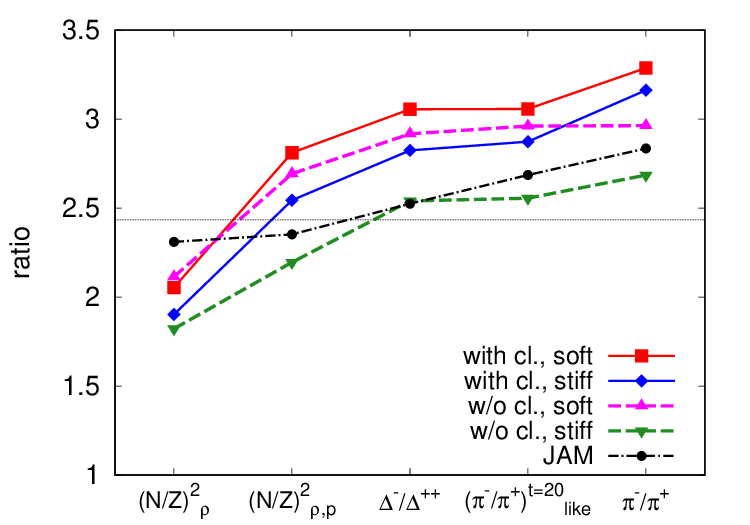}
\caption{The nucleon ratios $(N/Z)^2_{\rho}$ and $(N/Z)^2_{\rho, p}$
  [Eq.~(\ref{eq_NZ2})] in high-density region with and without
  high-momentum condition, respectively, the $\Delta^{-}/\Delta^{++}$
  production ratio [Eq.~(\ref{eq_Delta})], the pion-like ratio at
  $t=20$ fm/$c$, and the final $\pi^{-}/\pi^{+}$ ratio.  Each line
  connects the ratios for each of the five cases of calculation for
  central collisions of ${}^{132}\mathrm{Sn}+{}^{124}\mathrm{Sn}$ at
  300 MeV/nucleon. The horizontal line represents the
  $(N/Z)^2_{\rm sys}$ ratio of the total system.  The statistical
  uncertainties in the final $\pi^{-}/\pi^{+}$ ratio are smaller than
  0.02.}
\label{ratio_summary}
\end{center}
\end{figure}

To discuss the relation between the dynamics of nucleons and the final
pion ratio, we show in Fig.~\ref{ratio_summary} a summary of different
ratios in five calculations.

For the time-dependent nucleon ratio $[N(t)/Z(t)]^2$ shown in
Fig.~\ref{fig:np-ratio-squared}, we here define a representative
$(N/Z)^2$ ratio as
\begin{eqnarray}
\left( \frac{N}{Z} \right)^2
  =\frac{ \int_{0}^{\infty} N(t)^{2} dt}{\int_{0}^{\infty} Z(t)^{2} dt},
\label{eq_NZ2}
\end{eqnarray}
where $N(t)$ and $Z(t)$ indicate the numbers of neutrons and protons
as functions of time which satisfy the conditions described in the
previous subsection.  This ratio carries the information in the
compression stage because $N(t)$ and $Z(t)$ take large values only if
the system has a large high-density region $\rho>\rho_0$.  In the
first two columns of Fig.~\ref{ratio_summary}, the $(N/Z)^2_{\rho}$
and $(N/Z)^2_{\rho, p}$ ratios show the calculated representative
values when the nucleons are selected by the conditions (a) and (b),
respectively.  As we have already seen in the previous subsection, the
$(N/Z)^2$ increases by choosing the high momentum part of the phase
space.  The effect is stronger in the calculations with cluster
correlations.  This is understandable because, with cluster
correlations, many $\alpha$ clusters are formed which contain the same
number of neutrons and protons and have relatively low momentum per
nucleon, and therefore the remaining part of high-momentum nucleons
becomes neutron rich.  Another important point found here is that the
symmetry energy effect, namely the difference between stiff and soft
symmetry energies, is smaller when the cluster correlation is turned
on.  This is also reasonable because the cluster correlation forces
some neutrons and protons to move together, and therefore the
different forces acting on neutrons and protons are averaged out to
some degree.

As a representative value of the $\Delta$ production ratio, we show,
in the third column of Fig.~\ref{ratio_summary}, the
$\Delta^{-}/\Delta^{++}$ ratio of the total production numbers
\begin{eqnarray}
 \frac{\Delta^{-}}{\Delta^{++}}=
 \frac{ \int_{0}^{\infty}(nn \rightarrow p \Delta^{-})dt}{\int_{0}^{\infty} (pp \rightarrow n \Delta^{++})dt},
\label{eq_Delta}
\end{eqnarray}
where $(nn \rightarrow p \Delta^{-})$ and
$(pp \rightarrow n \Delta^{++})$ indicate the reaction rates of the
$\Delta$ production at each time shown in the upper panel of
Fig.~\ref{delta}.  We can see that the $\Delta^{-}/\Delta^{++}$ ratio
is different from the $(N/Z)^2_{\rho}$ ratio, while the
$(N/Z)^2_{\rho, p}$ ratio is almost equal to the
$\Delta^{-}/\Delta^{++}$ ratio.  These results are consistent with the
comparison of Fig.~\ref{fig:delta-production-ratio} and
Fig.~\ref{fig:np-ratio-squared}.

It might not be straightforward, in principle, how the $\Delta$
production ratio is related to the pion ratio because many of the
produced $\Delta$ resonances are absorbed by $N\Delta\rightarrow NN$
reactions as we have seen seen in Fig.~\ref{delta}.  However, we find
that the behavior of the $\Delta$ production ratio before
$t\approx 20$ fm/$c$ in Fig.~\ref{fig:delta-production-ratio} is
similar to the pion-like ratio in Fig.~\ref{pion-like_ratio}
calculated from the $\Delta$ resonances and pions that exist at each
time.  The ratio of numbers of $\Delta$ resonances at $t=20$ fm/$c$ is
$\Delta^{++}:\Delta^+:\Delta^0:\Delta^-\approx 1:1.62:2.49:3.58$ in
the lower panel of Fig.~\ref{delta} for the case of soft symmetry
energy with cluster correlations.  The $\Delta^-$ to $\Delta^{++}$
ratio is significantly larger than the $\Delta$ production ratio 
because the neutron-richness of the system influences the isospin
dependence of the $\Delta$ absorption rates\footnote{
  If the system were in chemical equilibrium with a temperature $T$
  and neutron and proton chemical potentials $\mu_n$ and $\mu_p$, one
  would expect the $\Delta$ multiplicity ratio might follow
  $\Delta^-/\Delta^{++} \simeq
  e^{3(\mu_n-\mu_p)/T}\simeq(N/Z)_{\rho,p}^3$
  for $(m_\Delta - \mu_{n,p})/T \gg 1$.  In our simulation, however,
  $\Delta^-/\Delta^{++}=3.58$ is not as large as $(N/Z)_{\rho,p}^3=4.7$.
 This implies that dynamical effects in the neutron and proton
 distributions are important and/or the full chemical equilibrium for
 $\Delta$ resonances is not achieved.
%
}.  
The pion-like ratio calculated from only these $\Delta$ resonances
is
$(\pi^-/\pi^+)'_{\text{like}}=(\Delta^{-}+\frac{1}{3}\Delta^{0})/(\Delta^{++}+\frac{1}{3}\Delta^{+})=$2.86
which happens to be similar to the $\Delta$ production ratio.  The
fourth column of Fig.~\ref{ratio_summary} shows the pion-like ratio at
$t=20$ fm/$c$ (see Fig.~\ref{pion-like_ratio}).  An interesting
observation is that the dependence of the pion-like ratio on the
symmetry energy and the cluster correlations is quite similar to that
of the $\Delta$ production ratio and therefore to that of
$(N/Z)^2_{\rho,p}$.  This suggests that the information on the
high-density nucleon dynamics remains in the pion-like ratio at $t=20$
fm/$c$, without being much influenced by the $\Delta$ absorption.

\begin{figure*}
\begin{center}
\includegraphics[width=8.5cm,height=6.5cm]{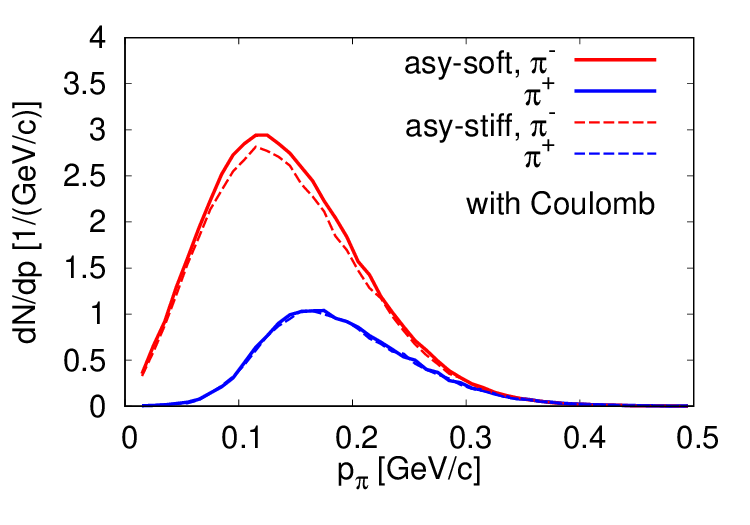}
\includegraphics[width=8.5cm,height=6.0cm]{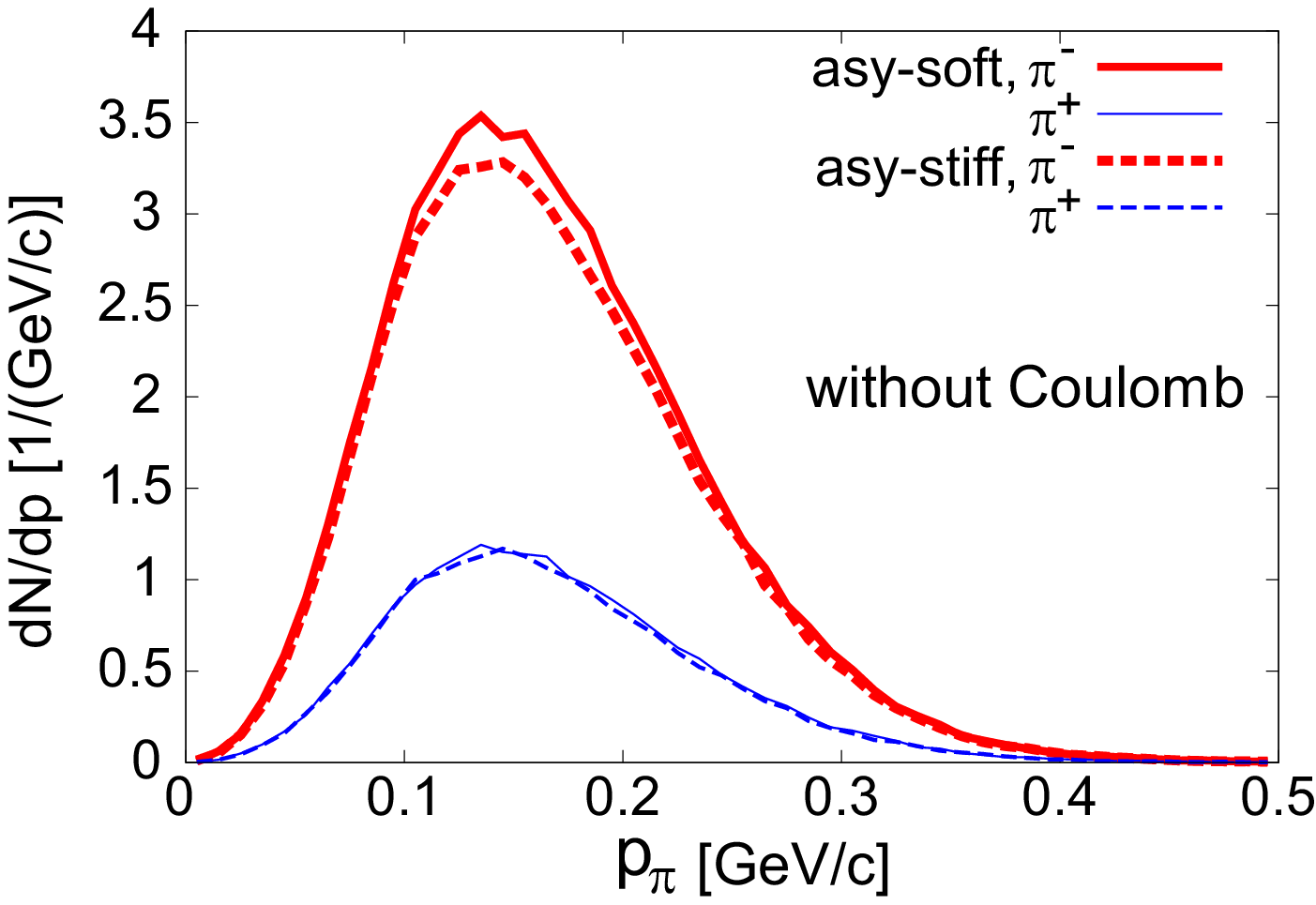}
\caption{Momentum distribution of $\pi^-$ and $\pi^+$ in the
  center-of-mass frame calculated with AMD+JAM with clusters for
  central collisions of ${}^{132}\mathrm{Sn}+{}^{124}\mathrm{Sn}$ at
  300 MeV/nucleon.  The left and right panels show the results with
  and without Coulomb force for pions, respectively.}
\label{fig:coulom1}
\end{center}
\end{figure*}
\begin{figure*}
\begin{center}
\includegraphics[width=8.5cm, height=6.5cm]{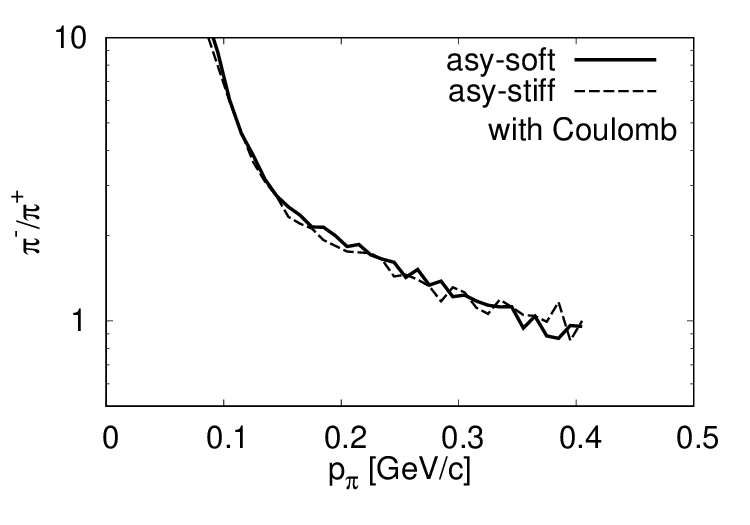}
\includegraphics[width=8.5cm, height=6.0cm]{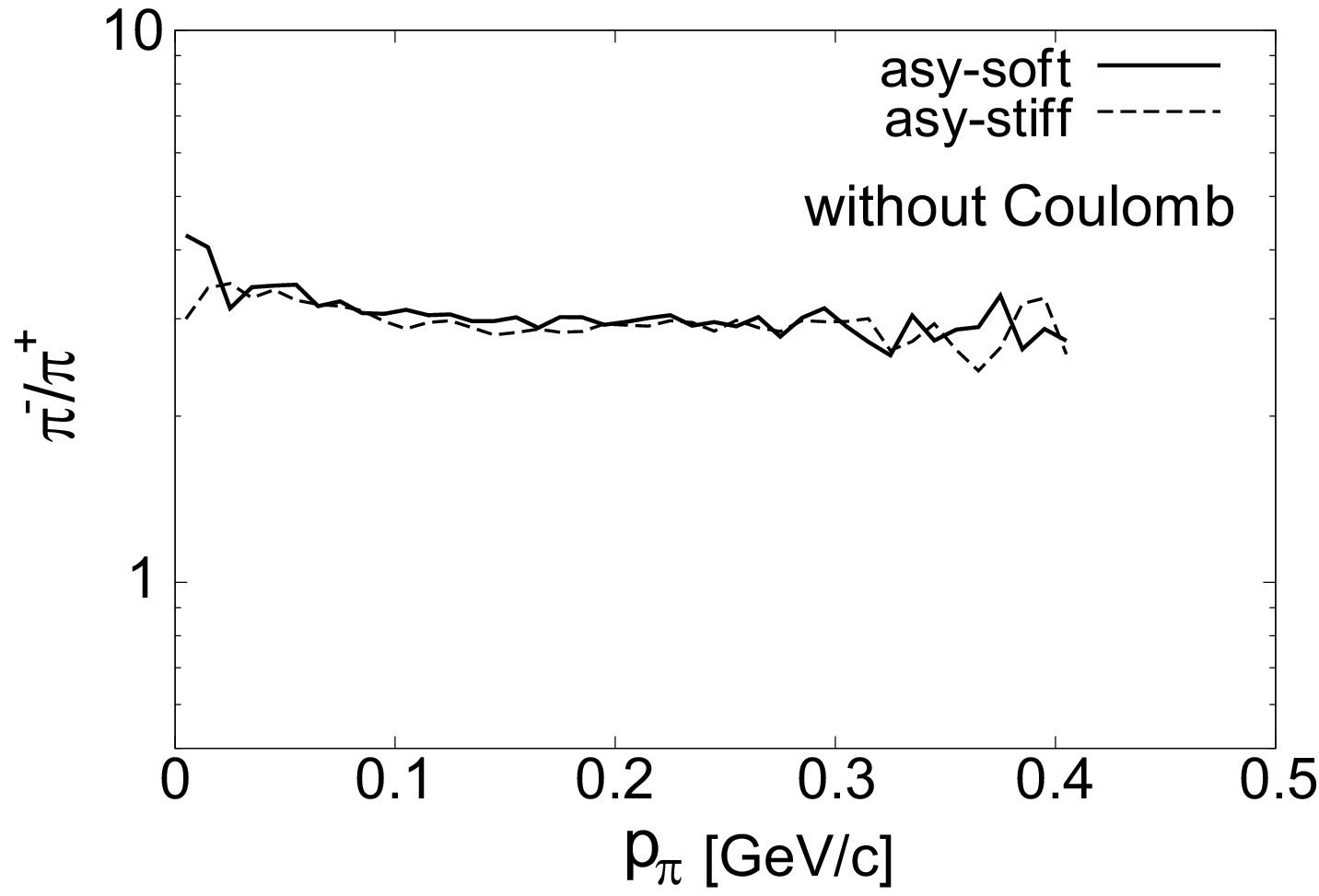}
\caption{The same as Fig.~\ref{fig:coulom1} but the $\pi^-/\pi^+$
  spectral ratio is shown.}
\label{fig:coulom2}
\end{center}
\end{figure*}

The pion-like particles have to go through the exterior region of the
expanding system.  The symmetry-energy effect on the nucleon $N/Z$
ratio in the exterior region should be opposite to that in the inner
region because the total numbers of neutrons and protons are (almost)
conserved.  As shown in the rightmost column of
Fig.~\ref{ratio_summary}, in the final stage of the reaction, the pion
ratio is modified to some degree from the pion-like ratio at $t=20$
fm/$c$ to the final $\pi^{-}/\pi^{+}$ ratio.  In each case of with and
without clusters, the symmetry energy effect at $t=20$ fm/$c$ is
reduced in the final ratio to about 70\% of the value at $t=20$
fm/$c$.  We also find the effect of clusters tends to raise the pion
ratio in the final stage, which is probably because the
interior/exterior part becomes less/more neutron rich when more
$\alpha$ clusters, with the same number of neutrons and protons
contained, are formed with relatively low velocities.

\subsection{Pion spectra}
Finally, we investigate the pion spectra.  Fig.~\ref{fig:coulom1}
shows the pion spectra with or without Coulomb force for charged
pions.  Coulomb force evidently changes the pion spectrum because
$\pi^{+}$ is accelerated and $\pi^{-}$ is decelerated.  The
$\pi^-/\pi^+$ spectral ratio shown in Fig.~\ref{fig:coulom2} can
become very large at low pion momenta due to the Coulomb effect.

The symmetry-energy effect, however, appears in our present results as
a simple normalization factor for the pion spectra and the spectral
ratio, and thus we cannot obtain information on the symmetry energy
effect more than found in the ratio of the total pion multiplicities.
This point does not agree with what has been found in the pBUU
calculation \cite{pBUU,Tsang-private} in which the symmetry-energy
effect is strong in the high momentum part of the spectra.

\section{Summary}
The mechanism of pion production was studied with a new approach by
combining two transport models AMD and JAM.  For central
${}^{132}\mathrm{Sn}+{}^{124}\mathrm{Sn}$ collisions at 300
MeV/nucleon, the production of $\Delta$ resonances and pions are
treated as perturbation.  Two different AMD calculations with and
without cluster correlations were performed, not only to investigate
the effect of clusters but also to study the correlations between the
nucleon dynamics and $\Delta$ and pion production.  We found that the
$\Delta^-/\Delta^{++}$ production ratio agrees very well with the
neutron-proton squared ratio $(N/Z)^2$ in the high-density and
high-momentum region of the one-body phase space.  
We also found that the $\Delta$ production ratio, and therefore
 $(N/Z)^2$, are directly reflected in the $\pi^-/\pi^+$ ratio.
The effect of the
high-density symmetry energy in the $\pi^-/\pi^+$ ratio is modified in
the final stage of the reaction, with a large part of the effect still
remaining, which is qualitatively similar to the case of BUU in the
literature \cite{Li}.

If the AMD calculations with and without clusters are regarded as two
different models, the present results show the value of the pion
ratio is model-dependent as well as the nucleon dynamics is.
Evidently models should be constrained by other observables such as
the multiplicities and the spectra of clusters in order to extract the
symmetry energy from the pion ratio.  It may also possible to find a
combination of observables that has less model dependence, such as by
taking the ratio of an observable from different reaction systems
\cite{BaoAnLi:2008}.  It is, nevertheless, preferable to clarify the
origin of the different predictions of different models by
investigating the dynamics in detail as has been done here. 

It is, of course, interesting and necessary to extend the present
study to other reaction systems with different neutron-proton
asymmetries and for different energies.  Such calculations are in
progress, and the results will be reported elsewhere.

\section*{Acknowledgments}
This work was supported by JSPS KAKENHI Grant Numbers 24105008, 
15K05079 and 15H06413.

\appendix
\section{Momentum dependent term\label{sec:mdcorr}}
The momentum-dependent term of the interaction energy density for a
Skyrme effective interaction can be written by using the phase-space
distribution function $f_\alpha(\bm{r},\bm{p})$
($\alpha=p\uparrow,p\downarrow,n\uparrow,n\downarrow$) as
\begin{multline}
\mathcal{E}_\tau(\bm{r})=\frac{1}{2}\sum_{\alpha\beta}
U^\tau_{\alpha\beta}
\int\frac{d\bm{p}_1}{(2\pi\hbar)^3}
\int\frac{d\bm{p}_2}{(2\pi\hbar)^3}
\\
(\bm{p}_1-\bm{p}_2)^2
f_\alpha(\bm{r},\bm{p}_1)f_\beta(\bm{r},\bm{p}_2).
\end{multline}
The coefficients $U^\tau_{\alpha\beta}$ are related to the
Skyrme parameters by
\begin{multline}
  U^\tau_{\alpha\beta}=
\tfrac{1}{4}t_1
\langle\alpha\beta|(1+x_1P_\sigma)|\alpha\beta-\beta\alpha\rangle
\\
+\tfrac{1}{4}t_2
\langle\alpha\beta|(1+x_2P_\sigma)|\alpha\beta+\beta\alpha\rangle.
\end{multline}
Employing the fact that the momentum dependence is quadratic, it is
possible to write
\begin{equation}
\mathcal{E}_\tau(\bm{r})
=\sum_{\alpha\beta}U^\tau_{\alpha\beta}
\bigl(\tau_\alpha(\bm{r})\rho_\beta(\bm{r})
-\bm{J}_\alpha(\bm{r})\cdot\bm{J}_\beta(\bm{r})\bigr),
\label{eq:ETskyrme}
\end{equation}
where several kinds of densities are defined by
\begin{align}
\rho_\alpha(\bm{r})&=\int\frac{d\bm{p}}{(2\pi\hbar)^3}f_\alpha(\bm{r},\bm{p}),\\
\bm{J}_\alpha(\bm{r})
&=\int\frac{d\bm{p}}{(2\pi\hbar)^3}
\bigl(\bm{p}-\bar{\bm{p}}(\bm{r})\bigr)f_\alpha(\bm{r},\bm{p}),\\
\tau_\alpha(\bm{r})
&=\int\frac{d\bm{p}}{(2\pi\hbar)^3}
\bigl(\bm{p}-\bar{\bm{p}}(\bm{r})\bigr)^2f_\alpha(\bm{r},\bm{p}).
\end{align}
Although $\bar{\bm{p}}(\bm{r})$ can be arbitrarily chosen for
Eq.~(\ref{eq:ETskyrme}) to hold, we will choose the local average
momentum
\begin{equation}
\bar{\bm{p}}(\bm{r})=\frac{1}{\sum_\alpha\rho_\alpha(\bm{r})}
\sum_\alpha\int\frac{d\bm{p}}{(2\pi\hbar)^3}
\bm{p}f_\alpha(\bm{r},\bm{p}),
\end{equation}
so that the term $\bm{J}_\alpha\cdot\bm{J}_\beta$ in Eq.~(\ref{eq:ETskyrme}) can
be neglected keeping the Galilei invariance.

For the distribution function of Eq.~(\ref{eq:amd-wigner}) in the case
of AMD, we can analytically perform the momentum integration to have
\begin{multline}
\tau_{\alpha}(\bm{r})
=\Bigl(\frac{2\nu}{\pi}\Bigr)^{\frac32}\sum_{j,k\in\alpha}
\bigl[(\bm{P}_{jk}-\bar{\bm{p}}(\bm{r}))^2
+3\hbar^2\nu\bigr]
\\\times
e^{-2\nu(\bm{r}-\bm{R}_{jk})^2}B_{jk}B^{-1}_{kj}.
\end{multline}
For the application to high energy collisions, we now modify the
quadratic momentum dependence to the same momentum dependence as
Ref.~\cite{GBDG} by modifying $\tau_\alpha(\bm{r})$ to
\begin{multline}
\tilde{\tau}_{\alpha}(\bm{r})
=\Bigl(\frac{2\nu}{\pi}\Bigr)^{\frac32}\sum_{j,k\in\alpha}
\Bigl[\frac{(\bm{P}_{jk}-\bar{\bm{p}}(\bm{r}))^2}
{1+(\bm{P}_{jk}-\bar{\bm{p}}(\bm{r}))^2/\Lambda^2}
+3\hbar^2\nu\Bigr]
\\\times
e^{-2\nu(\bm{r}-\bm{R}_{jk})^2}B_{jk}B^{-1}_{kj},
\end{multline}
and then use the modified momentum-dependent part of the interaction
energy density
\begin{equation}
\tilde{\mathcal{E}}_\tau(\bm{r})=
\sum_{\alpha\beta}U^\tau_{\alpha\beta}\tilde{\tau}_\alpha(\bm{r})\rho_\beta(\bm{r}).
\end{equation}
In the present work, we choose the parameter $\Lambda=395$ MeV/$c$ for
the momentum scale.

\section{Correlations to bind clusters \label{sec:coacc}} 

Many of light nuclei (Li, Be \textit{etc.}) have only one or a few
bound states which may be regarded as bound states of internal
clusters.  The quantum-mechanical probability of forming such a
nucleus is not consistent with the semiclassical phase space with
which it can be formed in the standard treatment of AMD.  Therefore,
for a better description, inter-cluster correlation is introduced as a
stochastic process of binding clusters.

The basic idea is to replace the radial component of the relative
momentum between clusters by zero if moderately separated clusters
($2.5<R_{\text{rel}}<7$ fm) are moving away from each other with a
small relative kinetic energy
[$\bm{R}_{\text{rel}}\cdot\bm{V}_{\text{rel}}>0$ and
$\frac12\mu(V_{\text{rel}\parallel}^2+0.25V_{\text{rel}\perp}^2)<7$
MeV].  In addition to these conditions, linking is allowed only if
each of the two clusters is one of the three closest clusters of the
other when the distance is measured by
$[(\bm{R}_{\text{rel}}/3\
\text{fm})^2+(\bm{V}_{\text{rel}}/0.25c)^2]^{1/2}$,
so that linking usually occurs in dilute environment.  Non-clustered
nucleons are treated here in the same way as clusters but two nucleons
are not allowed to be linked.  Two clusters also should not be linked
if they can form an $\alpha$ or lighter cluster due to the combination
of their spins and isospins.  It is possible that more than two
clusters are linked by this condition.  However, only in the case that
the mass number of the linked system is $\le 10$, the binding is
performed for the linked system by eliminating the radial velocities
of clusters in the center-of-mass frame of the linked system.

The energy conservation should be achieved by scaling the relative
radial momentum between the center-of-mass of the linked system and a
third cluster.  A reasonable way to choose a third cluster may be to
find a cluster which has participated in a collision that formed one
of the clusters in the linked system.  However, since we do not keep
the full history of collisions in our computation, we choose a cluster
that has the minimal value of
\begin{equation}
(r+7.5\ \text{fm})(1.2-\cos^2\theta)/
\min(\varepsilon_\parallel,\ 5\ \text{MeV})
\end{equation}
as the third cluster for energy conservation, where $r$ and
$\varepsilon_\parallel$ are the distance and the radial component of
the kinetic energy for the relative motion between the linked system
and the third cluster.  The factor with the angle $\theta$ between the
relative coordinate ($\bm{r}$) and velocity ($\bm{v}$) is introduced
so as to favor the case of $\bm{r}\parallel\bm{v}$.

\section{Test particles for the AMD phase-space distribution}\label{AppC}
Here we describe a method to generate test particles following the
one-body phase-space distribution function $f(\bm{r},\bm{p})$ given by
Eq.~(\ref{eq:amd-wigner}).  We generate $A$ test particles, where $A$
is the number of nucleons (for each spin-isospin state) in the system.

Let us first consider a distribution given by a sum of Gaussian distributions
\begin{equation}
g(\bm{r},\bm{p})=\sum_{i=1}^Ag_i(\bm{r},\bm{p}),
\end{equation}
with
\begin{equation}
g_i(\bm{r},\bm{p})=(2\alpha)^3
e^{-2\nu\alpha(\bm{r}-\bm{R}_i)^2
   -\alpha(\bm{p}-\bm{P}_i)^2/2\hbar^2\nu},
\end{equation}
where the `physical coordinates' \cite{AMD} are used as the centroids
$(\bm{R}_i,\bm{P}_i)$.  The case of $\alpha=1$ corresponds to the
usual wave-packet molecular dynamics (MD) without antisymmetrization.
A natural idea in MD is to sample a test particle from each Gaussian
distribution $g_i(\bm{r},\bm{p})$.  It should be noted that this MD
sampling introduces many-body correlations among test particles, which
is different from sampling $A$ test particles independently following
the total distribution $g(\bm{r},\bm{p})$.

Our aim here is to extend the MD sampling, with reasonable
correlations, for the total one-body distribution $f(\bm{r},\bm{p})$.
We may decompose $f(\bm{r},\bm{p})$ into $A$ terms as
\begin{equation}
f(\bm{r},\bm{p})=\sum_{i=1}^A
 \hat{f}(\bm{r},\bm{p})g_i(\bm{r},\bm{p})
\label{eq:amd-wigner-decomp}
\end{equation}
with $\hat{f}(\bm{r},\bm{p})=f(\bm{r},\bm{p})/g(\bm{r},\bm{p})$.  The
average number of test particles to be generated for each term should be
\begin{equation}
\bar{N}_i=\int\hat{f}(\bm{r},\bm{p})g_i(\bm{r},\bm{p})\frac{d\bm{r}d\bm{p}}{(2\pi\hbar)^3}
\approx\frac{1}{n}\sum_{j=1}^n\hat{f}(\bm{r}_j,\bm{p}_j)
\end{equation}
which is evaluated by sampling many points $(\bm{r}_j,\bm{p}_j)$
($j=1,2,\ldots,n$) from the Gaussian distribution
$g_i(\bm{r},\bm{p})$.  With these average numbers $\bar{N}_i$, the
actual numbers of test particles $N_i$, which should be integers, are
randomly determined in such a way that $\sum_{i=1}^AN_i=A$ and
$N_i=\mathop{\text{floor}}(\bar{N}_i)$ or
$\mathop{\text{floor}}(\bar{N}_i)+1$.

For each term of Eq.~(\ref{eq:amd-wigner-decomp}), $N_i$ test
particles should be sampled with the relative weight function
$\hat{f}(\bm{r},\bm{p})g_i(\bm{r},\bm{p})$, which is a straightforward
numerical procedure.  However, we may introduce additional
correlations among test particles by modifying $\hat{f}$ as
\begin{equation}
\hat{f}(\bm{r},\bm{p}):= (1-e^{-2\nu\beta(\bm{r}-\bm{r}_k)^2-\beta(\bm{p}-\bm{p}_k)^2/2\hbar^2\nu})\hat{f}(\bm{r},\bm{p})
\end{equation}
when a test particle $(\bm{r}_k,\bm{p}_k)$ is generated.  Test
particles are generated sequentially in a random order, and the
modification of $\hat{f}$ by the $k$-th test particle influences only
the test particles generated after $k$.

In the present work, we have chosen the parameters
$\alpha=\frac{2}{3}$, $\beta=2$ and $n=20$.

\end{document}